\documentclass[12pt]{JHEP}
\usepackage{epsfig}

\font\blackboard=msbm10 at 12pt
\font\blackboards=msbm7
\font\blackboardss=msbm5
\newfam\black
\textfont\black=\blackboard
\scriptfont\black=\blackboards
\scriptscriptfont\black=\blackboardss

\newcommand{\NP}{{\rm Nucl.\ Phys.\ }}

\newcommand{\PL}{{\rm Phys.\ Lett.\ }}
\newcommand{\PR}{{\rm Phys.\ Rev.\ }}

\def\math@note#1{\gdef\@eqnlabel{LAB: #1}}
\title{Perturbative diagrams in string field theory}
\author{Washington Taylor\\
{Center for Theoretical Physics} \\
{MIT, Bldg.  6-308} \\
{Cambridge, MA 02139, U.S.A.} \\
{\tt wati@mit.edu}}

\abstract{A general algorithm is presented which gives a closed-form
expression for an arbitrary perturbative diagram of cubic string field
theory at any loop order.  For any diagram, the resulting expression
is given by an integral of a function of several infinite matrices,
each built from a finite number of blocks containing the Neumann
coefficients of Witten's 3-string vertex.  The closed-form expression
for any diagram can be approximated by level truncation on oscillator
level, giving a computation involving finite size matrices.  Some
simple tree and loop diagrams are worked out as examples of this
approach.}

\preprint{MIT-CTP-3290, NI02015-MTH, hep-th/0207132}
\begin{document}

\baselineskip16pt
\parskip=4pt

\section{Perturbative diagrams in string field theory}
\label{sec:diagrams}

\subsection{Introduction}

String field theory is a formulation of string theory as a nonlocal
field theory of an infinite number of fields in space-time.  This
approach goes beyond the world-sheet formulation of string theory in
several ways.  First, it gives a systematic way of constructing
perturbative string amplitudes in terms of vertices and propagators;
this approach is in principle easier to generalize to higher loop
amplitudes than the world-sheet approach, which involves conformal
field theory on higher genus Riemann surfaces.  Second, string field
theory gives a nonperturbative off-shell formulation of string theory,
and can be used to address questions which go beyond string
perturbation theory.  Third, while all current formulations of string
field theory are formally described in a fixed string background,
fluctuations of the string background itself are naturally encoded in
the theory, so that string field theory is really a
background-independent theory.  Recent work using Witten's cubic open
string field theory \cite{Witten-SFT} to confirm Sen's conjectures
\cite{Sen-universality} regarding tachyon condensation has
demonstrated conclusively that string field theory accurately
describes some nonperturbative off-shell aspects of string physics
(for reviews see \cite{Ohmori,deSmet,abgkm}).  Furthermore, the
existence of a nontrivial vacuum solution of string field theory
demonstrates the background independence of the theory
\cite{ks-open,Sen-Zwiebach,Moeller-Taylor}.

In this paper we focus on the first of the points mentioned in the
previous paragraph, namely the construction of arbitrary perturbative
string amplitudes using string field theory.  Constructing higher-loop
amplitudes in string theory using covariant world-sheet methods
becomes difficult when the Riemann surface involved has genus greater
than one.  This is because the calculation of a generic string
amplitude involves an integral over the moduli space of the
appropriate Riemann surface, and it is difficult to define an
appropriate measure on the moduli space for higher-genus surfaces.  On
the other hand, a string field theory such as Witten's cubic open
string field theory gives a straightforward construction, in
principle, of any higher-loop amplitude.  The amplitude can be
expressed in standard field-theoretic language in terms of the cubic
vertices and the propagators for the infinite number of space-time
fields.  The difficulty with doing such a calculation explicitly is
that the infinite number of fields have complicated cubic interactions
described by the Witten 3-string vertex.

Formal arguments have demonstrated that in a particular gauge
(Feynman-Siegel gauge), Witten's cubic bosonic open string field
theory gives rise to a diagrammatic expansion which precisely covers
the moduli space of Riemann surfaces of arbitrary genus with at least
one boundary and with an arbitrary number of open string punctures on
the boundaries \cite{gmw,Zwiebach-proof}.  These results show
that this string field theory must precisely reproduce all
perturbative on-shell string amplitudes given by the conformal field
theory of the bosonic string in 26 dimensions.  Explicit computations
of perturbative amplitudes using string field theory, however, have
only been done for the tree level 4-point function and the one-loop
2-point function \cite{Giddings,Sloan,Samuel-off,fgst}.  These computations were
done using the conformal mapping method, and required an
explicit mapping between the modular parameters associated with the
string field theory parameterization of moduli space and the standard
parameterization of conformal field theory.  Such mappings lead to
complicated formulae even for tree-level and genus one calculations,
and are unknown beyond genus one.

In this paper, we take a different approach to reproducing
perturbative string amplitudes from string field theory.  Rather than
trying to relate string field theory calculations to conformal field
theory, we simply proceed directly to evaluate perturbative amplitudes
using the oscillator representation of the vertices and propagators in
Witten's cubic string field theory.  Since the vertex and propagator
can be written completely in terms of squeezed states, we can give a
closed-form expression for any perturbative amplitude, even at higher
loop order.  The only complication is the appearance of
infinite-dimensional matrices in the final expression for each
diagram.  These matrices are built from simple blocks, however, which
can be expressed in terms of the Neumann coefficient of Witten's
3-string vertex.  While we do not yet have the technology to
analytically evaluate amplitudes written in terms of these matrices,
we can evaluate any desired amplitude numerically to a high
degree of precision using truncation on the level of oscillators which
are included.  This truncation method is more powerful than the method of
level truncation on fields used to address the tachyon
condensation problem in
\cite{ks-open,Sen-Zwiebach,WT-SFT,Moeller-Taylor}, since rather than
having to include a 
number of fields which grows exponentially in the level, we simply
need to evaluate the determinant of some matrices whose size grows
linearly in the truncation level.  

In Section 1.2 we give a brief review of Witten's cubic open string
field theory to fix notation.  In Section 1.3 we present an algorithm
for computing a closed-form expression for any perturbative string
amplitude.  Section 2 contains a number of examples of tree- and
loop-level diagrams: at tree level we analyze the 4-point and 5-point
functions for the tachyon, at one loop we describe the tadpole, and at
two loops we describe the 0-point function.  We perform numerical
evaluations of some of these amplitudes, showing that the procedure of
level convergence on oscillator level (amounting to a sort of UV
cutoff from the open string point of view) converges quickly.

\subsection{Witten's cubic string field theory}

In this subsection we give a brief synopsis of Witten's cubic string
field theory \cite{Witten-SFT}.  For further background on open
string field theory see \cite{Witten-SFT,lpp,Gaberdiel-Zwiebach}.

The degrees of freedom of Witten's cubic string field theory are
encoded in a string field $\Phi$.  $\Phi$ can be considered as a
functional $\Phi[x (\sigma); c (\sigma), b (\sigma)]$ of the matter,
ghost, and antighost configuration of the string.  $\Phi$ can also be
thought of as living in the string Fock space ${\cal H}$ spanned by
states produced by acting with a finite number of oscillators on the
string vacuum
\begin{equation}
\Phi = 
\int d^{26} p \; \left[
\phi (p)\; | 0; p \rangle + A_\mu (p) \; \alpha^\mu_{-1} | 0; p
\rangle + \cdots \right]\,.
\end{equation}
In this expression, $| 0 ; p\rangle$ is the ghost number 1 vacuum at
momentum $p$, which is annihilated by matter, antighost, and ghost
modes $a_n, b_n, c_n$ with $n \geq 1$.  In this paper we use matter
oscillators with canonical commutation relations $[a_n, a_{-m}] =
\delta_{nm}$.

The action of Witten's theory can be written
\begin{equation}
S = -\frac{1}{2}\langle V_2 | \Phi, Q \Phi \rangle
- \frac{g}{3}  \langle V_3 | \Phi, \Phi, \Phi \rangle
\label{eq:action}
\end{equation}
where $| V_2 \rangle \in {\cal H}^2, | V_3 \rangle \in {\cal H}^3$.
Explicit oscillator representations of $| V_2 \rangle, | V_3 \rangle$
are given by
\begin{eqnarray}
| V_2 \rangle & = &  \int d^{26} p \;
\exp\left(-a^{(1)}_{-n} C_{nm} a^{(2)}_{-m}
-c^{(1)}_{-n} C_{nm}  b^{ (2)}_{-m}-c^{(2)}_{-n} C_{nm} 
b^{(1)}_{-m}\right)  \times
\label{eq:v2}\\
& &\hspace{1in}
(c_0^{(1)} + c_0^{(2)})
\left(| 0; p \rangle \otimes | 0; -p \rangle \right) \nonumber\\
| V_3 \rangle & = &  \int d^{26} p^{(1)}\, d^{26} p^{(2)} \;
\exp\left(- \frac{1}{2} a^{(i)}_{-n} {N}^{ij}_{nm} a^{(j)}_{-m}
-  a^{(i)}_{-n} {N}^{ij}_{n0} p^{(j)}
- \frac{1}{2} p^{(i)} {N}^{ij}_{00} p^{(j)}
-c^{(i)}_{-n} {X}^{ij}_{nm} b^{(j)}_{-m}\right)\times \nonumber\\
& &\hspace{1in}
(c_0^{(1)}c_0^{(2)}c_0^{(3)})
\left(| 0;  p^{(1)} \rangle \otimes | 0; p^{(2)} \rangle
\otimes | 0; p^{(3)} =-p^{(1)}- p^{(2)} \rangle) \right)  \label{eq:v3}
\end{eqnarray}
where $Q$ is the open string BRST operator,
\begin{equation}
C_{nm} = \delta_{nm} (-1)^n\,,
\end{equation}
and where $N^{ij}_{nm}, X^{ij}_{nm}$ are Neumann coefficients for
which exact expressions are given in
\cite{Gross-Jevicki-12,cst,Samuel,Ohta}.  The values of these
coefficients are tabulated for $n + m < 10$ in \cite{WT-SFT}.  (Note
that with the conventions we are using here, the Neumann coefficients
$N^{ij}_{nm}$ in that reference should be rescaled by  factors of
$-\sqrt{nm}$.  We have also removed from $| V_3 \rangle$ an overall
numerical factor of $k = (3 \sqrt{3}/2)^3$.  Note also that
in (\ref{eq:v2}, \ref{eq:v3}), all summations over indices $n, m$ are
taken over $n, m \geq 1$, except the last summation over $m$, which is
taken over $m \geq 0$ so that the mode $b_0$ is included.)

The action (\ref{eq:action}) has an enormous gauge symmetry.  A
convenient choice of gauge for perturbative calculations is
Feynman-Siegel gauge, where $b_0 | \Phi \rangle = 0$.  In this gauge,
all fields associated with states having a $c_0$ acting on the vacuum
are taken to vanish.  This simplifies the above vertices in that we
can ignore all ghost 0 modes $c_0, b_0$.  Furthermore, the BRST
operator in this gauge is simply 
\begin{equation}
Q =  c_0 L_0 = c_0 (p^2 + N^{(m)} + N^{(g)} -1)\,.
\end{equation}
All calculations in this paper are done in Feynman-Siegel gauge.
In this gauge the propagator is given by (dropping the factor of $c_0$)
\begin{equation}
\frac{1}{L_0}  = \int_0^\infty dT \;e^{-TL_0}\,.
\label{eq:}
\end{equation}

\subsection{Algorithm for computing amplitudes}
\label{sec:algorithm}

We now present a general algorithm which gives a closed-form
expression for any perturbative open string diagram in Feynman-Siegel
gauge.  A number of examples are worked out explicitly in the following
section.

Consider any diagram with $v$ cubic vertices and $e$ internal edges.
Label the half-edges in the diagram with integers from 1 through $3v$,
with labels 1, 2, 3 for the half-edges connected to vertex 1, labels
4, 5, 6 for the half-edges connected to vertex 2, and so forth.
Denote the half-edges associated with the $k$th internal edge by $i_k,
j_k$, so that the first edge connects the half-edges $i_1, j_1$, etc.
The $v$ 3-string vertices can be represented in the $3v$-fold tensor
product of the single string Fock space through
\begin{equation}
| V \rangle = | V_3 \rangle_{123} \otimes
| V_3 \rangle_{456} \otimes
\cdots  \otimes| V_3 \rangle_{(3v-2) (3v-1) (3v)}  \in
{\cal H}^{3v}\,.
\label{eq:v}
\end{equation}
The propagator in Feynman-Siegel gauge for the $e$ internal edges can
be written by acting with half of each propagator on each
associated half-edge, giving an operator on ${\cal H}^{3v}$
\begin{equation}
P =\int \prod_{k = 1}^{e} d T_k \; e^{-\frac{1}{2}T_k 
(2 p_k^2 + N_{i_k} + N_{j_k}-2)}\,.
\end{equation}
The constraint that the half-edges $i_k, j_k$ are connected can be
simply imposed by contracting with the dual state
\begin{eqnarray}
\langle D | & = &
\left(\prod_{k = 1}^{e}
  \int d^{26} p \; \langle p_{i_k} = p | \otimes \langle  p_{j_k} = -p |
\right)\times\label{eq:dual-state}\\
 &  & \hspace{1in}
\exp \left(\sum_{ k = 1}^{e}  -a^{(i_k)}_n C_{nm} a^{(j_k)}_m
-c^{(i_k)}_n C_{nm}  b^{(j_k)}_m-c^{(j_k)}_n C_{nm} 
b^{( i_k)}_m\right)\,.\nonumber
\end{eqnarray}
We drop all factors of $c_0, b_0$ from (\ref{eq:v}) and
(\ref{eq:dual-state}) as they automatically cancel for calculations in
Feynman-Siegel gauge.
The full amplitude for the diagram under consideration is now given
by an integral over internal (loop) momenta
\begin{equation}
{\cal A} = \int \prod_{i = 1}^{1 + e-v}  d^{26} q_i \;
\langle D | P | V \rangle\,.
\label{eq:amplitude}
\end{equation}
This amplitude is a state in ${\cal H}^{3v-2e}$, and can be contracted
with any external string states to get any particular amplitude
associated with the relevant diagram.

Since all the pieces of (\ref{eq:amplitude}) are given in terms of
exponentials of quadratic expressions in the oscillators, we can give
a closed form expression for any diagram using standard squeezed state
techniques.  It is convenient to first compute $P | V \rangle$, which
is given by
\begin{eqnarray}
P | V \rangle  & = & \int \prod_{k = 1}^{e}  (dT_k e^{T_k(1-p_k^2)}) 
\times\;\\
 & & \hspace{0.5in}
\exp \left( - \frac{1}{2} a^{(i)}_{-n} \tilde{N}^{ij}_{nm} a^{(j)}_{-m}
- a^{(i)}_{-n} \tilde{N}^{ij}_{n0} p^{(j)}
- \frac{1}{2} p^{(i)} {N}^{ij}_{00} p^{(j)}
-c^{(i)}_{-n} \tilde{X}^{ij}_{nm} b^{(j)}_{-m}\right)
\left( \prod_{i = 1}^{3v} | p_i \rangle \right) , \nonumber
\end{eqnarray}
where $\tilde{N}$ is a block-diagonal matrix consisting of 3 by 3
blocks of infinite matrices $N^{rs}$ of Neumann coefficients
associated with each 3-string vertex, multiplied by an exponential
factor for the appropriate propagator(s).  For a pair of internal
indices $i_k, j_l$ at a common vertex, we have
\begin{equation}
\tilde{N}^{i_k j_l}_{nm} = \hat{N}^{ij}_{nm} (e^{-T_k}, e^{-T_l}) \equiv
 e^{-n T_k/2} N^{ij}_{nm} e^{-m T_l/2},
\label{eq:nti}
\end{equation}
where $i, j$ are equal to $i_k, j_l$ modulo 3.  For indices at a given
vertex associated with external edges, the exponential factors are
dropped, which can be expressed by replacing the relevant argument(s)
of $\hat{N}$ by 1.  The matrix $\tilde{X}$ is similarly constructed
from the ghost Neumann coefficients $X^{ij}_{mn}$ of the Witten
vertex.  We can now remove all oscillators associated with internal
edges using the matter equation \cite{Kostelecky-Potting}
\begin{eqnarray}
\lefteqn{\langle 0 | \exp \left( -{\frac{1}{2}a \cdot S \cdot a}\right)
\exp \left({-\mu \cdot  a^{\dagger} -\frac{1}{2}
a^{\dagger}
\cdot \tilde{N} \cdot a^{\dagger}}\right) | 0 \rangle
} \\
 &  &\hspace{1in}=  \frac{1}{ \det (1-S \tilde{N})^{1/2}} 
\exp \left( -\frac{1}{2} \mu \cdot (1-S \tilde{N})^{-1} S \cdot\mu\right)
\nonumber
\end{eqnarray}
and the associated ghost equation
\begin{eqnarray}
\lefteqn{\langle 0 | \exp \left(-c \cdot S \cdot b\right)
\exp \left({-\lambda_c \cdot b^{\dagger} - c^{\dagger} \cdot \lambda_b  -
c^{\dagger}
\cdot \tilde{X} \cdot b^{\dagger}}\right) | 0 \rangle} \\
 & 
& \hspace{2in}=\det (1+S \tilde{X})
\exp \left( -\lambda_c \cdot (1-S \tilde{X})^{-1} S
\cdot\lambda_b\right)\,. \nonumber
\end{eqnarray}
This gives us an expression for the diagram of interest in terms of an
integral over internal momenta and moduli of the form
\begin{eqnarray}
{\cal A} & = & 
\int \left(\prod_{i = 1}^{1 + e-v}  d^{26} q_i \right) \;
\left(\prod_{k = 1}^{e}  dT_k \; e^{T_k}\right)\;
\frac{\det (1+S \tilde{X})}{ \det (1-S \tilde{N})^{13}} 
\label{eq:amplitude-full}\\
 & & \hspace{2in} \times
\exp \left({ -\frac{1}{2}(a^{\dagger}, p_i) \cdot Q \cdot  (a^{\dagger},
 p_i) -
c^{\dagger} \cdot R \cdot b^{\dagger}}\right) 
\left(\prod_{i = 1}^{3v-2e}  | p_i \rangle \right)\,. \nonumber
\end{eqnarray}
In this expression, $\tilde{X}$ and $\tilde{N}$ are as above, and
depend on the modular parameters $T_i$.  The determinants are
evaluated in the $2e \times \infty$ dimensional space associated with
internal edges.  The 0 components of these matrices associated with
momentum are not included in the determinant; all momentum factors are
left explicitly in the exponential.  The matrix $S$ is simply a
permutation matrix expressing which half-edges are connected by
propagators, tensored with the infinite matrix $C$.  The oscillators
$a^{\dagger}, b^{\dagger}, c^{\dagger}$ are the raising operators
associated with matter, antighost and ghost fields of the external
edges; the notation $(a^{\dagger}, p_i)$ indicates a vector including
the matter raising operators of external edges, as well as all
external and internal momenta.  The quadratic forms $Q$ and $R$ are
matrices depending on the modular parameters $T_i$; these quadratic
forms take the schematic form $N + N(1-S \tilde{N})^{-1} S N$ and $X +
X (1-S \tilde{X})^{-1} SX$.  Since all integrals over the internal
momenta $q_i$ are Gaussian, these can be readily performed, leaving an
integral over the modular parameters associated with the edge lengths $T_i$.

This completes the algorithm for constructing a closed-form expression
for any perturbative string diagram using Witten's cubic string field
theory.  A number of examples are worked out explicitly in the
following section, which should help to clarify details which have
been suppressed in this concise description.  We conclude this
section by discussing some general features of this approach.

\subsection{Comments on the algorithm}

\subsubsection{Level truncation}

Because the matrices involved in these diagrams are infinite, it is
difficult to analytically evaluate the integrals in
(\ref{eq:amplitude-full}), even for simple diagrams.  It is easy,
however, to truncate these matrices at a finite oscillator level,
giving an expression which can be computed and numerically integrated,
even up to high oscillator levels.  From the point of view of the open
string, this truncation essentially amounts to imposing a UV cutoff on
string theory.  This regularization breaks many of the desirable
features of string theory, such as the large gauge symmetry and
duality symmetries.  Nonetheless, as we will describe in the following
section, this truncation allows us to accurately approximate many
on-shell and off-shell quantities of interest.

\subsubsection{Equivalent formulations}
\label{sec:equivalent}

Instead of placing all the vertices on the right in
(\ref{eq:amplitude}), we could put some vertices on the left and some
on the right.  An example of such a calculation is mentioned below in
connection with (\ref{eq:a4-alternate}).  While the resulting
expressions cannot  be written as easily for a general diagram, this is
often a convenient trick to simplify the evaluation of some particular
amplitude at a fixed level of truncation.  For tree-level diagrams, it
is always possible to place half the vertices on the left and the
other half on the right, leading to a reduction in the size of the
matrices by a factor of 2.  A similar procedure is possible for loop
diagrams, although each fundamental loop with an odd number of
vertices requires an additional block in the matrices.

\subsubsection{Analytic evaluation}

We do not yet have tools adequate to perform a direct analytic
evaluation of any string amplitudes constructed in this fashion.
Recent work \cite{diagonalization}
on the diagonalization of the Neumann coefficient 
matrices $N^{rs}$, however, gives hope that progress may be possible
towards an analytic understanding of the determinants involved in
(\ref{eq:amplitude-full}).  In particular, it would be interesting
to ascertain whether a similar diagonalization could be carried out
for the Neumann coefficient matrices after factors of $e^{-n T_r/2}$
are multiplied into the matrices on each side.
Results in this direction might lead to  new methods for analytically
computing on-shell and off-shell string amplitudes.

\section{Examples}

In this section we present several examples of the algorithm described
in Section~\ref{sec:algorithm}.

\subsection{Tree-level  4-tachyon amplitude at $p = 0$}

For a first simple example, consider the off-shell tree-level 4-point
function of the tachyon, where all external edges are taken to have
momentum $p = 0$.  The relevant diagram is depicted in
Figure~\ref{f:diagram1}, where all incoming momenta are taken to
vanish.  Because our external states are all tachyons, we can drop all
oscillators associated with external edges, and we only need to
concern ourselves with the oscillators associated with the two halves
of the single internal edge.  Since all momenta vanish, we can drop
all momenta from the calculation.
\FIGURE{
\epsfig{file=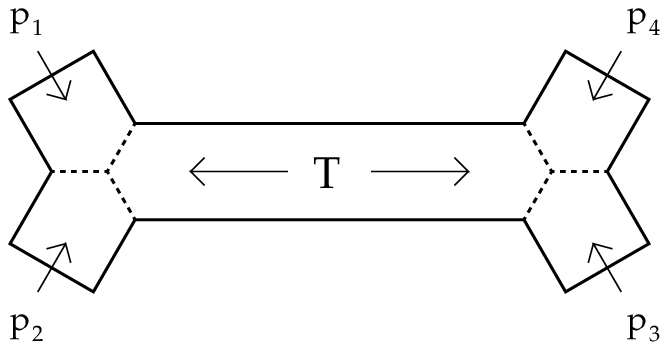,width=15cm}
\caption{\footnotesize Tree-level 4-point function}
\label{f:diagram1}
}

For this diagram, the permutation matrix $S$ connecting the half-edges
is given by
\begin{equation}
S =\left(\begin{array}{cc}
0 & C\\
C & 0
\end{array} \right)\,.
\end{equation}
The matrix $\tilde{N}$ can be written in terms of $2 \times 2$
infinite blocks as
\begin{equation}
\tilde{N} (T) =\left(\begin{array}{cc}
\hat{N}^{11}_{nm} (e^{-T}, e^{-T}) &  0\\
0 &\hat{N}^{11}_{nm} (e^{-T}, e^{-T})
\end{array} \right),
\label{eq:n1}
\end{equation}
where
\begin{equation}
\hat{N}^{11}_{nm} (x, y) = x^{n/2} N^{11}_{nm} y^{m/2}\,.
\end{equation}
Similarly, we have
\begin{equation}
\tilde{X}(T) =\left(\begin{array}{cc}
\hat{X}^{11}_{nm} (e^{-T}, e^{-T})  &  0\\
0 & \hat{ X}^{11}_{nm} (e^{-T}, e^{-T})
\end{array} \right)\,.
\label{eq:x1}
\end{equation}
The off-shell $p = 0$ tree-level 4-tachyon amplitude is then given by
\begin{equation}
{\cal A}_4 = \int_0^\infty dT \; e^{T} \;
\frac{\det (1+S \tilde{X}(T))}{\det (1-S \tilde{N} (T))^{13}} \,.
\label{eq:4ta}
\end{equation}
Note that this amplitude can also be written
\begin{equation}
{\cal A}_4 = \int_0^1 \frac{dx}{x^2} 
\frac{\det \left(1-[C\hat{X}^{11} (x, x)]^2\right)}{\det
\left(1-[C\hat{N}^{11} (x, x)]^2\right)} .
\label{eq:a4-alternate}
\end{equation}
This form of the amplitude can be derived from (\ref{eq:4ta}), or can
be constructed directly by modifying the general algorithm so that one
vertex is included on the left in (\ref{eq:amplitude}) and the other
on the right, as mentioned in \ref{sec:equivalent}.  This form of the
amplitude is useful because it involves smaller matrices which are
easier to compute in level truncation.

If we remove by hand from the amplitude (\ref{eq:4ta}) the
contribution coming from the intermediate tachyon state, we get the
coefficient $c_4$ of $\phi^4$ in the effective action for the $p = 0$
tachyon field $\phi$ when all other string fields are integrated out.
An integral expression for this coefficient was given in
\cite{ks-exact}, and evaluated numerically to an accuracy of 1\%,
giving $c_4 \approx -1.75 \pm 0.02$.  We have repeated this
calculation to a higher degree of precision, with the result
\begin{equation}
c_4 \approx -1.742 \pm 0.001\,.
\label{eq:c4-best}
\end{equation}
The coefficient $c_4$ was computed
using level truncation including all intermediate fields up to level 4
in \cite{ks-open} and including fields up to level 20 in
\cite{WT-SFT}.  The level 20 approximation gives $c^{[20]}_4 =
-1.684$, within 5\% of the exact result.

Including the numerical factors associated with the
symmetry of the diagram in Figure~\ref{f:diagram1}, we have 
\begin{equation}
c_4 = \frac{9}{2} \int_0^\infty dT \; e^{T} \;
\left[
\frac{\det (1+S \tilde{X}(T))}{\det (1-S \tilde{N} (T))^{13}} -1 \right] \,.
\label{eq:c4-exact}
\end{equation}
It is instructive to see how the level truncation in oscillator level
works as a systematic approximation scheme for (\ref{eq:c4-exact}).
The simplest level truncation involves dropping all oscillators other
than $a_1, a_{-1}$.  We then have the finite-size matrices
\begin{equation}
S =\left(\begin{array}{cc}
0 & -1\\
-1 & 0
\end{array} \right),
\label{eq:s11}
\end{equation}
\begin{equation}
\tilde{N} (T) =\left(\begin{array}{cc}
\frac{5}{27}  e^{-T} &  0\\
0 & \frac{5}{27}  e^{-T}
\end{array} \right),
\label{eq:n11}
\end{equation}
\begin{equation}
\tilde{X}(T) =\left(\begin{array}{cc}
-\frac{11}{27}  e^{-T} &  0\\
0 & -\frac{11}{27}  e^{-T}
\end{array} \right),
\label{eq:x11}
\end{equation}
where we have used $N^{11}_{11} = 5/27, X^{11}_{11} = -11/27$ (note
again that the sign convention used here is opposite to that used in
\cite{WT-SFT}).   
Evaluating the determinants and replacing $x = e^{-T}$ gives
\begin{equation}
c_4^{(1)} = -\frac{9}{2} \int_0^1 \frac{dx}{x^2} 
\left[\frac{1-\frac{121}{729} x^2}{ (1-\frac{25}{729} x^2)^{13}}  -1 \right]\,.
\label{eq:c4-l1}
\end{equation}
This approximation to $c_4$ includes contributions from an infinite
number of intermediate fields---namely, all fields associated with
states produced by acting with $b_{-1} c_{-1}$ and $(a_{-1} \cdot
a_{-1})^k$ on the ground state.  Performing a power series expansion
in $x$, we have
\begin{equation}
c_4 \approx - \frac{9}{2} \int_0^1 dx \;
\left[\frac{68}{243}  + \frac{650}{19683}  x^2 + \cdots \right]\,.
\label{eq:c4e}
\end{equation}
Including only the first term, we have
\begin{equation}
c_4 \approx -\frac{34}{27}  \approx -1.26\,.
\label{eq:c41}
\end{equation}
This term represents all contributions from intermediate states at
level 2; all such states can be constructed using oscillators of mode
number 1.  This number agrees with the calculation using level
truncation on the level of intermediate states in
\cite{ks-open,WT-SFT}.  Including the second term in (\ref{eq:c4e})
shifts $c_4$ to $\approx -1.30879$.  This shift arises from the
intermediate fields associated with the states $(a_{-1} \cdot
a_{-1})^2 | 0 \rangle$ and $(a_{-1} \cdot a_{-1}) b_{-1} c_{-1} | 0
\rangle$, and again agrees with previous calculations.  Integrating
(\ref{eq:c4-l1}) gives
\begin{equation}
c_4^{(1)} =
-\frac{540022938263719272104656592800485}{673673135232465394630232928944128}
+ \frac{14196819}{10485760}  \ln \frac{11}{16}  \approx -1.30891\,.
\end{equation}
This shows that the contribution to $c_4$ from high level fields
composed completely of level 1 oscillators is quite small.

We can get successfully better approximations to (\ref{eq:c4-exact})
by truncating (\ref{eq:n1}, \ref{eq:x1}) at higher oscillator level
$L$.  Because this calculation simply involves numerically integrating
a determinant of a symmetric $L \times L$ matrix, this calculation is
significantly more efficient than using level truncation on states and
separately summing over all intermediate states in the diagram, as
done in \cite{ks-open,WT-SFT}.  Numerical results for the approximation
$c_4^{(L)}$ truncated at oscillator level $L \leq 100$ are shown in
Figure~\ref{f:graph-c4}.  At level 100, we have $c_4^{(100)} =
-1.734$, within 1\% of the value (\ref{eq:c4-best}).  Studying the
large $L$ behavior of $c_4^{(L)}$ suggests that the error introduced
by truncating at level $L$ decreases with a leading term of order
$1/L$.  A least-squares fit for the last 50 terms gives
\begin{equation}
c_4^{(L)} \approx -1.742 + \frac{0.80}{L} +\cdots
\end{equation}
in complete agreement with (\ref{eq:c4-best}).  It would be
interesting to find an analytic proof that contributions including
oscillators at
level $L$ (and no higher-level oscillators) decrease at the rate $L^{-2}$.
\FIGURE{
\epsfig{file=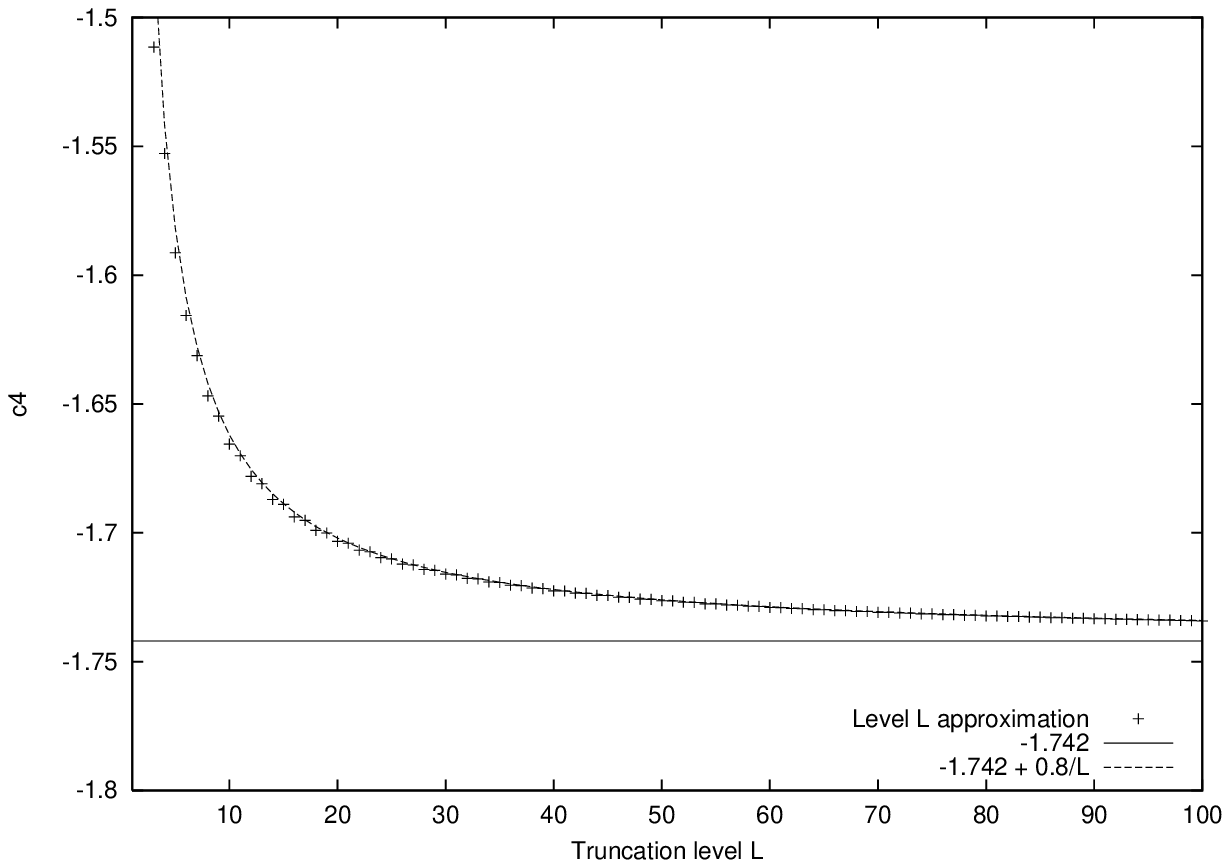,width=15cm}
\caption{\footnotesize Level-truncated approximations to $c_4$}
\label{f:graph-c4}
}

It is interesting to see how the integrand in (\ref{eq:c4-exact})
approaches its exact form as $L\rightarrow\infty$.  Analytic formulae
for this integrand were calculated by Sloan \cite{Sloan} and by Samuel
\cite{Samuel-off}, following Giddings' analysis of the on-shell
Veneziano amplitude \cite{Giddings} using the conformal mapping
approach.  We summarize some of this work in the Appendix, where Samuel's
analytic formula for the integrand is related to the integrand of
(\ref{eq:4ta}) through the relation (\ref{eq:analytic-integrand}).  In
Figure~\ref{f:graph-c4i}, we have graphed the integrand (without the
subtraction of the divergent tachyon term) at various cutoff levels
$L$ and compared to the exact formula (\ref{eq:analytic-integrand}).
The integrand converges very rapidly (apparently exponentially
rapidly) except near $x = 1$.  This rapid convergence away from $x =
1$ is natural, as the cutoff at level $L$ gets all terms in a power
series expansion of the integrand in $x$ correct up to order $x^{L +
1}$.  The parameter $x$ goes to 1 when the length of the intermediate
propagator goes to 0, so that level truncation here is playing the
role of a UV cutoff.  \FIGURE{ \epsfig{file=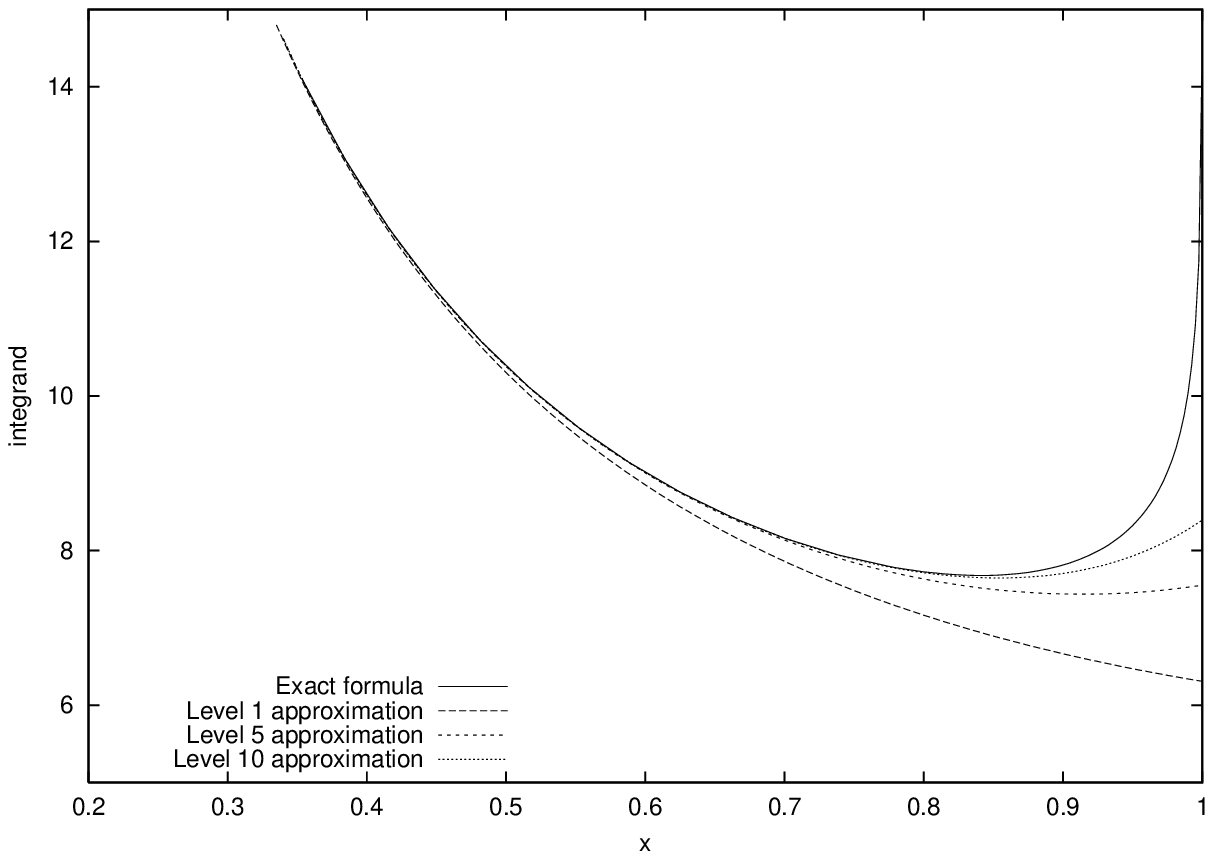,width=15cm}
\caption{\footnotesize Momentum-independent 4-tachyon integrand}
\label{f:graph-c4i}
}

\subsection{Tree-level 4-tachyon  amplitude ($p_i \neq 0$)}

Now let us consider the tachyon 4-point function with arbitrary
external momenta.  The on-shell tachyon 4-point function was computed
from string field theory in \cite{Giddings} using the conformal
mapping method, and shown to agree with the Veneziano amplitude.
Later, the off-shell tachyon 4-point function was computed using
similar methods in \cite{Sloan,Samuel-off}.  An expression for the
amplitude in terms of infinite matrices was also derived in
\cite{Samuel-off} using the oscillator approach, but it was not shown that
this expression agreed with the analytic formula arising from the
conformal mapping approach.  In this subsection we summarize the
results of using our algorithm to directly compute the Veneziano
amplitude and its off-shell generalization.  The relationship between
these results and those of \cite{Giddings,Sloan,Samuel-off} are discussed
in the Appendix.

In the 4-point tachyon diagram shown in Figure~\ref{f:diagram1},
momentum conservation ensures $p_1 + p_2 +p_3 + p_4 = 0$.  
The on-shell Veneziano amplitude associated with this diagram is
(setting $\alpha' = 1$)
\begin{equation}
{\cal A}^{[V]}_4 (p_1, p_2, p_3, p_4) =
B ( -s-1, -t-1)
\label{eq:Veneziano}
\end{equation}
where $B (u, v)$ is the Euler beta function, and $s, t$ are the
Mandelstam variables
\begin{equation}
s = -(p_1 + p_2)^2, \;\;\;\;\;
t = -(p_2 + p_3)^2, \;\;\;\;\;
\end{equation}
The Euler beta function has the integral representation
\begin{equation}
B (u, v) = \int_0^1 d\xi \;  \xi^{u-1} (1-\xi)^{v-1}\,.
\label{eq:integral-beta}
\end{equation}
This integral representation of the Veneziano amplitude is convergent
when $u, v > 0$, so that $s, t < -1$.  For positive real $s, t$ it is
necessary to perform an analytic continuation to make sense of
(\ref{eq:integral-beta}) away from the poles at positive integers.
The discovery of the Veneziano amplitude (\ref{eq:Veneziano})  marked
the beginning of string theory \cite{Veneziano}.  We will now show how
this amplitude can be reproduced numerically from string field theory
using our general approach to amplitude calculation.

The 4-point tachyon amplitude with arbitrary external momenta takes
the general form (\ref{eq:amplitude-full}), where contraction with
external tachyon states means that all oscillators $a^{\dagger},
b^{\dagger}, c^{\dagger}$ in the exponent can be dropped.  The
momentum-independent part of the
integrand in this case is identical to that of (\ref{eq:4ta}), which
is just the general amplitude with all momenta vanishing.  The
momentum terms appearing in the initial amplitude
(\ref{eq:amplitude}), before removing internal oscillators, are given
by
\begin{eqnarray}
 &  &\exp \left(-\frac{1}{2} N^{11}_{00}[p_1^2 + p_2^2 + p_3^2 + p_4^2 + 2
(p_1 + p_2)^2]\right) \; \times\\
& &\hspace{1in}
\exp \left(-T (p_1 + p_2)^2\right)
\exp \left(-p_i D^{ij}_n a^{(j)}_{-n} \right) \nonumber
\end{eqnarray}
where
\begin{equation}
D_n = \left(\begin{array}{cc}
\hat{N}^{12}_{0n} (1, x) - \hat{N}^{22}_{0n} (1, x) \;\;\; & 0\\
\;\;\;\hat{N}^{32}_{0n} (1, x) - \hat{N}^{22}_{0n} (1, x) \;\;\; & 0\\
0 &\;\;\;\hat{N}^{23}_{0n} (1, x) - \hat{N}^{33}_{0n}(1, x) \;\;\;\\
 0 & \;\;\;\hat{N}^{13}_{0n} (1, x) -\hat{N}^{33}_{0n}(1, x) \;\;\;
\end{array} \right)
\end{equation}
After contracting internal oscillators, this gives us a total
amplitude
\begin{equation}
{\cal A}_4 (p_1, p_2, p_3, p_4) =I (s, t) + I (t, s)
\label{eq:4tt}
\end{equation}
where
\begin{equation}
I (s, t) 
=
\frac{3^9}{2^{12}} 
 \int_0^\infty dT \; e^{T} \;
\frac{\det (1+S \tilde{X}(T))}{\det (1-S \tilde{N} (T))^{13}} 
\exp \left(-\frac{1}{2} p_iQ_{ij}p_j\right)
\label{eq:4tap}
\end{equation}
with
\begin{equation}
Q_{ij} = N^{11}_{00}\left(\begin{array}{cccc}
2 &1 & 0 & 0\\
1& 2 & 0 & 0\\
0 & 0 &  2 &1\\
0 & 0 &1 & 2
\end{array} \right)  +T
\left(\begin{array}{cccc}
1 &1 & 0 & 0\\
1& 1 & 0 & 0\\
0 & 0 & 1 &1\\
0 & 0 &1 & 1
\end{array} \right)  + D \frac{1}{1-S \tilde{N}}  SD^T\,.
\label{eq:4tq}
\end{equation}
Equations (\ref{eq:4tap}, \ref{eq:4tq}) give a complete description of
the off-shell generalization of the Veneziano amplitude
(\ref{eq:Veneziano}).
We have included in (\ref{eq:4tap}) the constant factor
needed for agreement with (\ref{eq:Veneziano}).  The two
terms in (\ref{eq:4tt}) come from integration over the regions $\xi <
1/2$, $\xi > 1/2$, as discussed in the Appendix.

Let us now consider the level truncation approximations of
(\ref{eq:4tap}, \ref{eq:4tq}).  At level 1, the matrices $S, \tilde{N},
\tilde{X}$ in the integrand of (\ref{eq:4tap}) are given by
(\ref{eq:s11}-\ref{eq:x11}).  Using the values $N^{11}_{00} = \ln
(27/16), N^{12}_{01} = -N^{21}_{01} =2 \sqrt{2}/3 \sqrt{3}$, and their
cyclic relatives, we have
\begin{equation}
Q^{(1)}_{ij} =\left(\begin{array}{cccc}
\;\;\;\ln\frac{3^6}{2^8 x}  + \frac{40x^2}{ 729-25x^2} \;\;\;&
\;\;\;\ln\frac{3^3}{2^4 x}  - \frac{40x^2}{ 729-25x^2}\;\;\;&
- \frac{216 x^2}{ 729-25x^2}&
 \frac{216 x^2}{ 729-25x^2}\\
\ln\frac{3^3}{2^4 x}  - \frac{40x^2}{ 729-25x^2}&
\ln\frac{3^6}{2^8 x}  + \frac{40x^2}{ 729-25x^2}&
 \frac{216 x^2}{ 729-25x^2}&
- \frac{216 x^2}{ 729-25x^2}\\
- \frac{216 x^2}{ 729-25x^2}&
 \frac{216 x^2}{ 729-25x^2}&
\;\;\;\ln\frac{3^6}{2^8 x}  + \frac{40x^2}{ 729-25x^2}\;\;\;&
\;\;\;\ln\frac{3^3}{2^4 x}  - \frac{40x^2}{ 729-25x^2} \;\;\;\\
 \frac{216 x^2}{ 729-25x^2}&
- \frac{216 x^2}{ 729-25x^2}&
\ln\frac{3^3}{2^4 x}  - \frac{40x^2}{ 729-25x^2}&
\ln\frac{3^6}{2^8 x}  + \frac{40x^2}{ 729-25x^2}
\end{array} \right)
\end{equation}
The level 1 approximation to (\ref{eq:4tap}) is thus given by
\begin{equation}
{\cal A}^{(1)}_4 (p_1, p_2, p_3, p_4) = I^{(1)} (s, t) + I^{(1)} (t, s)
\label{eq:4t1}
\end{equation}
where
\begin{equation}
I (s, t) =
\frac{3^9}{2^{12}} 
\int_0^1 \frac{dx}{x^2} 
\;\frac{1-\frac{121}{729} x^2}{ (1-\frac{25}{729} x^2)^{13}}   \;
\exp \left(-\frac{1}{2} p_iQ^{(1)}_{ij}p_j\right)\,.
\label{eq:Veneziano-integral}
\end{equation}
\FIGURE{
\epsfig{file=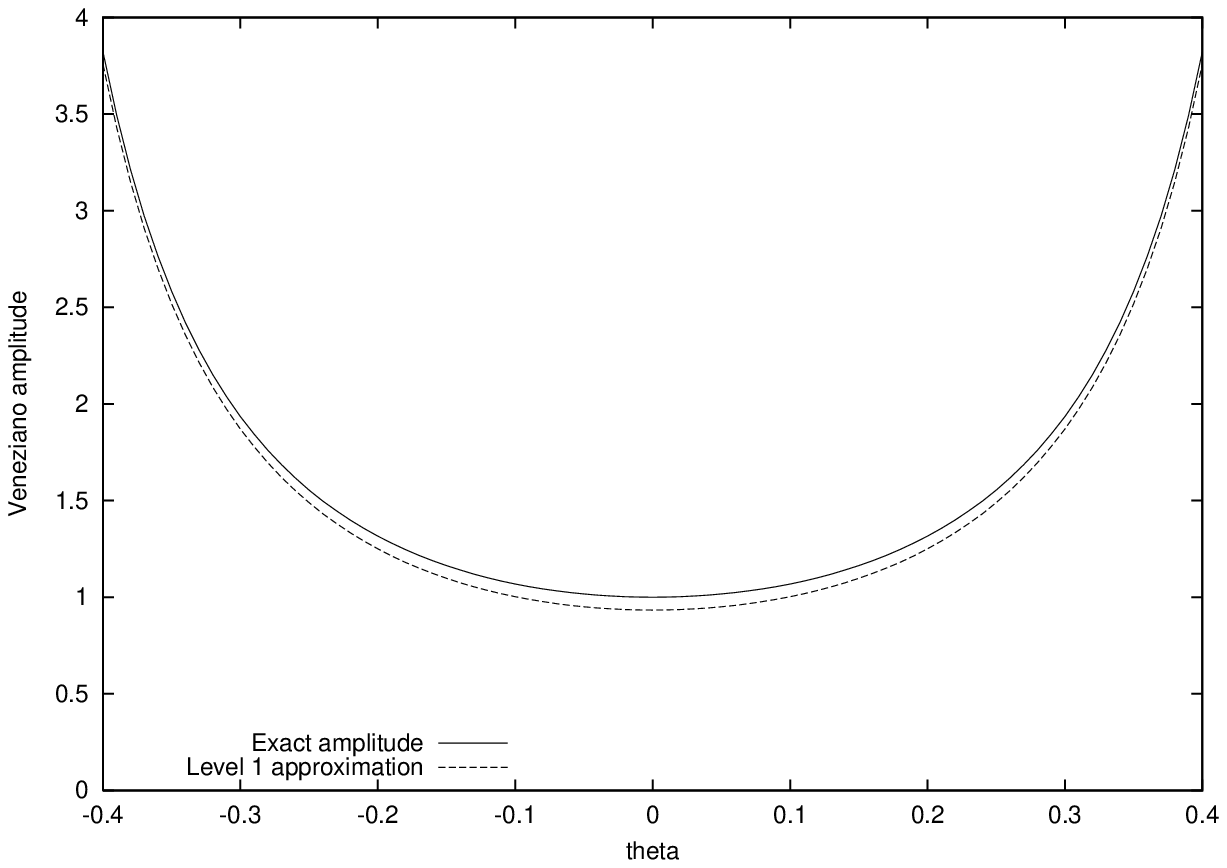,width=15cm}
\caption{\footnotesize Veneziano amplitude at $s = -2-2 \sin \theta, t
= -2 + 2 \sin \theta$}
\label{f:Veneziano1}
}

While the integral in (\ref{eq:4t1}) and the analogous integrals at
higher levels of truncation are not analytically tractable, such
integrals can be readily approximated numerically or by truncating a
power series expansion for the integrand.  To see how accurate the
approximation (\ref{eq:4t1}) is, we can numerically compute the
integral and compare to the Veneziano amplitude for particular values
of the momenta.  Let us consider in particular the case where  $u =
-(p_1 + p_3)^2 = 0$.  In this case,  when all particles are on-shell
tachyons we have $s + t = -4$.  We expect to have a convergent
integral when $- 3 < s, t < -1$.  In this regime we can choose momenta
\begin{eqnarray}
p_1 = (0, 1, 0, 0, \ldots, 0) & \hspace{1in} & 
p_2 = (0, \sin \theta,  \cos \theta, 0, \ldots, 0) \\
p_3 = (0, -1, 0, 0, \ldots, 0) & \hspace{1in} & 
p_4 = (0, -\sin \theta,  -\cos \theta, 0, \ldots, 0) 
\end{eqnarray}
parameterized by $\theta$, so that
\begin{equation}
s = -2-2 \sin \theta, \;\;\;\;\; t = -2 + 2 \sin \theta\,.
\end{equation}
We expect to have a convergent expression for the amplitude when
$-\pi/6 < \theta < \pi/6 \approx 0.5236$.  In
Figure~\ref{f:Veneziano1} we have graphed the Veneziano amplitude in
the range $| \theta | \leq 0.4$, and we have plotted the numerical
values for the level 1 approximation (\ref{eq:4t1}).

\FIGURE{
\epsfig{file=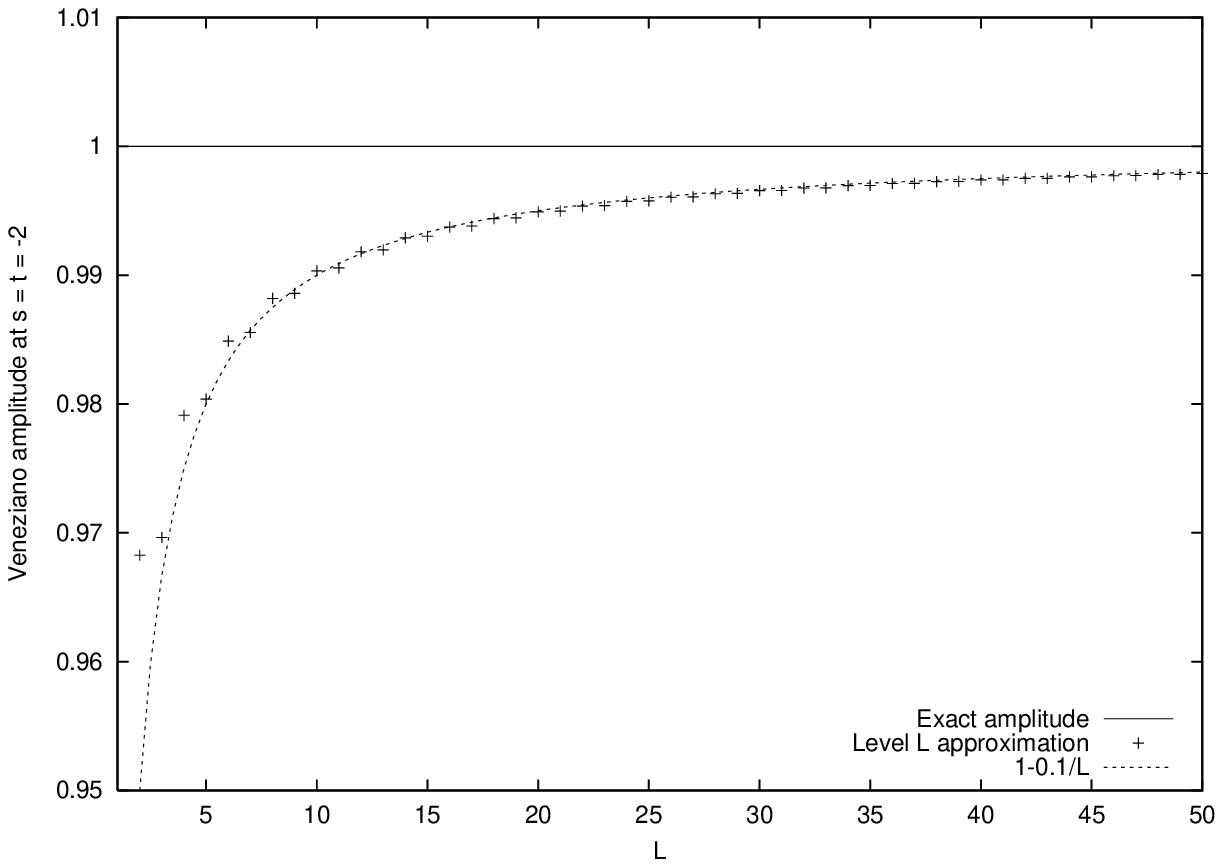,width=15cm}
\caption{\footnotesize Veneziano amplitude at  $s = t = -2$}
\label{f:Veneziano2}
}

The level 1 approximation to the Veneziano amplitude shown in
Figure~\ref{f:Veneziano1} is within 10\% of the correct result at
$\theta = 0$, and within 5\% of the correct result at $\theta = 0.4$.
In Figure~\ref{f:Veneziano2} we have graphed the successive
approximations to Veneziano at $\theta = 0$ ($s = t = -2$), up to
level $L = 50$.  At $\theta = 0$ the exact Veneziano amplitude is $B
(1, 1) = 1$.  The approximation at level 50 is 
\begin{equation}
{\cal A}^{(50)} (s = t = -2) = 0.9979
\end{equation}
within about $0.2\%$ of the exact answer.  Just as for the coefficient
$c_4$, it seems that the successive level approximations have errors
which decrease as $1/L$.  A least-squares fit for the last 25 terms
gives
\begin{equation}
{\cal A}^{(L)} (s = t = -2) \approx 0.99993  -0.10/L
\end{equation}

Although the particular values of $p_i$ for which we have computed
here the approximation to Veneziano are on-shell, the same analysis
can be done for off-shell amplitudes.  The discussion in the previous
subsection amounts to doing this in the case $p_i = 0$.  We have found
that in both the on-shell and off-shell cases the numerical
approximations to the amplitudes given by truncation on oscillator
mode level converge in a similar fashion, although the constant
controlling the size of the error is smaller for the on-shell calculation.

We have explicitly described here the 4-point function for external
tachyon states.  This could be generalized in a straightforward
fashion to include arbitrary external states by including the
oscillators $a^{(i)}$ on the external edges.  In Section
\ref{sec:1l1p} we do such a calculation for the one-loop one-point
function, which has only one external string.

\subsection{Tree-level $N$-point function}

It is easy to generalize the discussion of the previous subsection  to
tree-level $N$-point functions, although as $N$ increases the details
become more complicated.  We work out one further example here at tree
level, that of the 5-tachyon amplitude at $p = 0$.
\vspace{0.02in}

\noindent
{\bf 5-point function for 0-momentum tachyon}

\FIGURE{
\epsfig{file=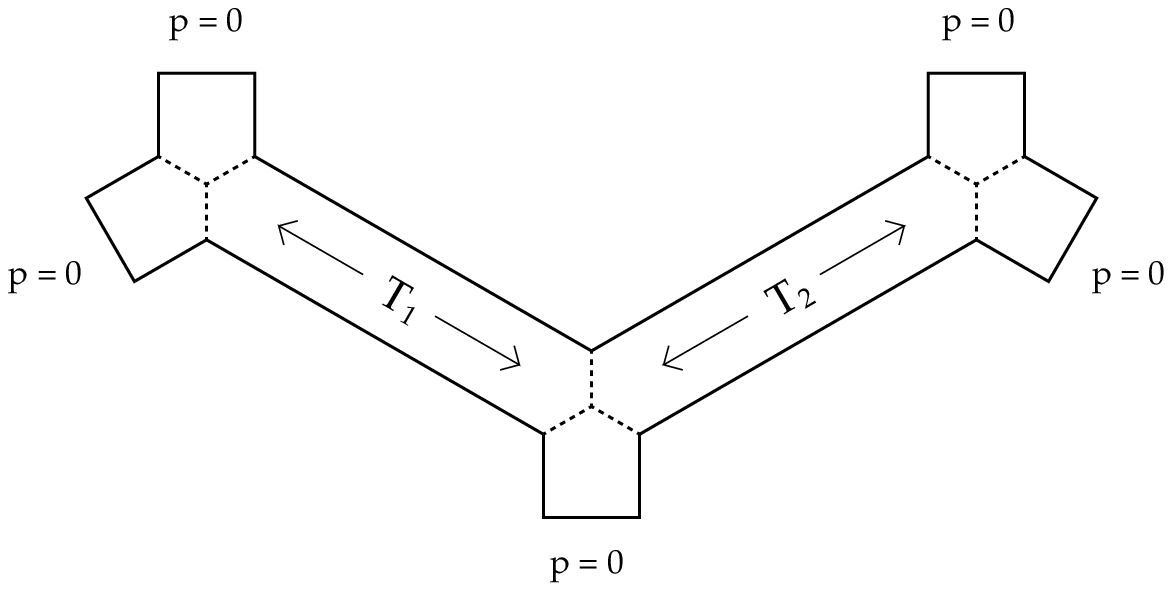,width=15cm}
\caption{\footnotesize Tree-level $p = 0$ 5-tachyon amplitude}
\label{f:t5}
}

The 5-point function for the tachyon at $p = 0$ comes from the diagram
in Figure~\ref{f:t5}.  This amplitude is given by
\begin{equation}
{\cal A}_5 = \int \frac{dx}{x^2}  \frac{dy}{y^2}  \;
\frac{\det (1+S \tilde{X})}{\det (1-S \tilde{N} )^{13}} 
\label{eq:a5}
\end{equation}
where
\begin{equation}
x = e^{-T_1}, \;\;\;\;\;
y = e^{-T_2}, 
\end{equation}
\begin{equation}
S = \left(\begin{array}{cccc}
0 & C & 0 & 0\\
C & 0 & 0 & 0\\
0 & 0 & 0 & C\\
0 & 0 & C & 0
\end{array}\right),
\end{equation}
\begin{equation}
\tilde{N} = \left(\begin{array}{cccc}
\hat{N}^{11}(x, x) & 0 & 0 & 0\\
0 & \hat{N}^{22}(x, x)&\hat{N}^{23}(x, y) & 0\\
0 & \hat{N}^{32}(y, x)&\hat{N}^{33}(y, y) & 0\\
0 & 0 & 0 & \hat{N}^{11}(y, y)
\end{array}\right),
\label{eq:n5}
\end{equation}
and  an identical expression to (\ref{eq:n5}) gives $\tilde{X}$, where
all appearances of $\hat{N}$ are replaced by $\hat{X}$.  Writing the
integrand in (\ref{eq:a5}) as $F (x, y)$, the coefficient $c_5$ in the
tachyon effective potential is given by
\begin{equation}
c_5 = 27
\int \frac{dx}{x^2}  \frac{dy}{y^2}  \;
\left[ F (x, y) -F (x, 0) -F (0, y) +F (0, 0) \right],
\label{eq:c5}
\end{equation}
where the subtractions have the effect of removing the terms
associated with intermediate open string tachyon states.  

\FIGURE{
\epsfig{file=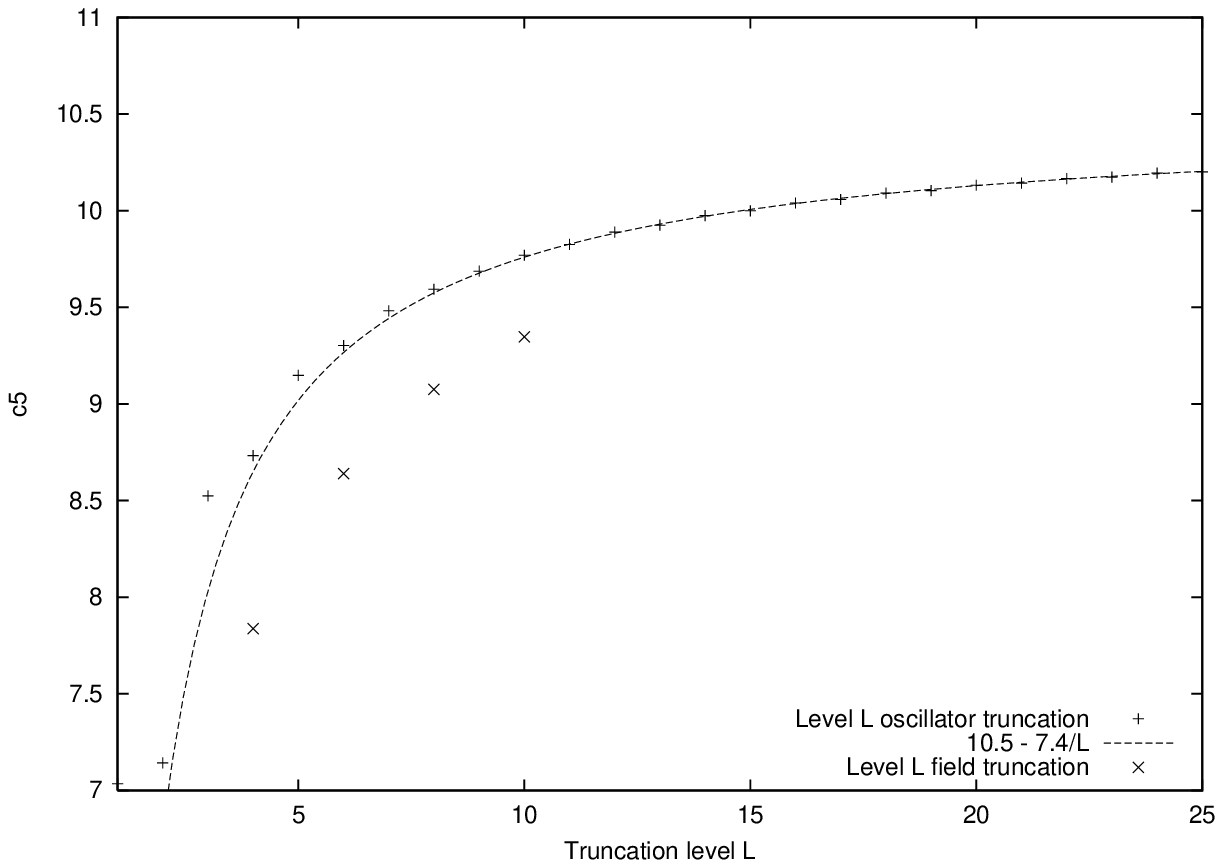,width=15cm}
\caption{\footnotesize Approximations to $c_5$}
\label{f:c5}
}

No exact expression is known for the coefficient $c_5$.
Approximate
values of $c_5$ were computed in \cite{Moeller-Taylor} using level
truncation on fields, up to level 10.  At level 10, we found
$c_5^{[10]} = 9.35$.  Approximations computed using
(\ref{eq:c5}) are graphed in Figure~\ref{f:c5} using oscillators up to
level 25.  The results of \cite{Moeller-Taylor} using level truncation
on fields are included for comparison.  At level 25 we have
$c_5^{(25)} = 10.20$.  Although we have less data in this case, it
seems that again the error goes as $1/L$.
A least-squares fit on the last 10 data points gives
\begin{equation}
c_5^{(L)} \approx 10.5 -7.4/L\,.
\end{equation}

\subsection{One-loop 1-point function}
\label{sec:1l1p}

Consider now the one-loop one-point function shown in
Figure~\ref{f:1l1p}.  By momentum conservation the external momentum
is 0, while we must integrate over the internal momentum $q$.  In this
example we will include all external oscillators, so that this
one-point function will be a state $|{\cal S} \rangle$ in the Fock
space.  To compute the tadpole for any particular field, the
appropriate state should be contracted with $|{\cal S} \rangle$.
\FIGURE{
\epsfig{file=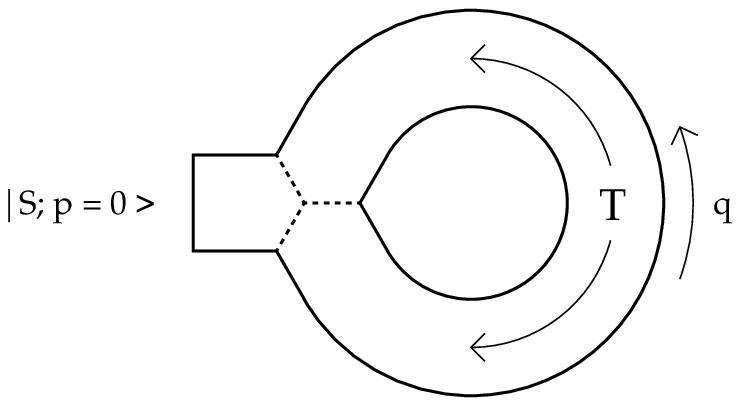,width=15cm}
\caption{\footnotesize One-loop one-point diagram}
\label{f:1l1p}
}

Using the general formalism described in \ref{sec:algorithm}, we have
\begin{eqnarray}
|{\cal S} \rangle  & = & 
\int_0^1 \frac{dx}{x^2}  \;
\int d^{26} q \;
\frac{\det (1+S \tilde{X})}{ \det (1-S \tilde{N})^{13}}\label{eq:sq}\\
 &  & \hspace{1in}
\times\exp \left( -\frac{1}{2} a_{-n}Q_{nm} a_{-m} - a_{-n} Q_n q -
\frac{1}{2} q^2 Q - c_{-n} R_{nm} b_{-m}
\right)| 0; 0 \rangle \nonumber
\end{eqnarray}
where
\begin{equation}
S =\left(\begin{array}{cc}
0 & C\\
C & 0
\end{array} \right),\;\;\;\;\;
\tilde{N} =\left(\begin{array}{cc}
\hat{N}^{22} (x, x) & \hat{N}^{23}(x, x)\\
\hat{N}^{32}(x, x) &  \hat{N}^{33}(x, x)
\end{array} \right),\;\;\;\;\;
\tilde{X} =\left(\begin{array}{cc}
\hat{X}^{22}(x, x) & \hat{X}^{23}(x, x)\\
\hat{X}^{32}(x, x) &  \hat{X}^{33}(x, x)
\end{array} \right)
\end{equation}
and
\begin{eqnarray}
Q_{nm}& = & N^{11}_{nm} +
\left(\begin{array}{c}
\hat{N}^{12}_{n \cdot}(1, x) \\
\hat{N}^{13}_{n \cdot}(1, x) 
\end{array} \right)^T
\frac{1}{1-S \tilde{N}}  S
\left(\begin{array}{c}
\hat{N}^{21}_{\cdot m}(x, 1) \\
\hat{N}^{31}_{ \cdot m}(x, 1) 
\end{array} \right) \nonumber\\
Q_n & = & N^{12}_{n0} -N^{13}_{n0} +
\left(\begin{array}{c}
\hat{N}^{12}_{n \cdot}(1, x) \\
\hat{N}^{13}_{n \cdot}(1, x) 
\end{array} \right)^T
\frac{1}{1-S \tilde{N}}  S
\left(\begin{array}{c}
\hat{N}^{22}_{ \cdot 0}(x, 1) -\hat{N}^{23}_{ \cdot 0}  (x, 1)\\
\hat{N}^{32}_{ \cdot 0}(x, 1) -\hat{N}^{33}_{ \cdot 0}  (x, 1)
\end{array} \right)\\
Q& = & N^{11}_{00} +
\left(\begin{array}{c}
\hat{N}^{22}_{0 \cdot}(1, x) -\hat{N}^{32}_{0 \cdot}  (1, x)\\
\hat{N}^{23}_{0 \cdot}(1, x) -\hat{N}^{33}_{0 \cdot}  (1, x)
\end{array} \right)^T
\frac{1}{1-S \tilde{N}}  S
\left(\begin{array}{c}
\hat{N}^{22}_{ \cdot 0}(x, 1) -\hat{N}^{23}_{ \cdot 0}  (x, 1)\\
\hat{N}^{32}_{ \cdot 0}(x, 1) -\hat{N}^{33}_{ \cdot 0}  (x, 1)
\end{array} \right) \nonumber\\
R_{nm}& = & X^{11}_{nm} +
\left(\begin{array}{c}
\hat{X}^{12}_{n \cdot}(1, x) \\
\hat{X}^{13}_{n \cdot}(1, x) 
\end{array} \right)^T
\frac{1}{1-S \tilde{X}}  S
\left(\begin{array}{c}
\hat{X}^{21}_{\cdot m}(x, 1) \\
\hat{X}^{31}_{ \cdot m}(x, 1) 
\end{array} \right) \nonumber
\end{eqnarray}

Integrating (\ref{eq:sq}) over the internal momentum $q$ gives (up to
a constant)
\begin{equation}
|{\cal S} \rangle =
\int_0^1 \frac{dx}{x^2}  \;
\frac{\det (1+S \tilde{X})}{ \det (1-S \tilde{N})^{13} Q^{13}}
\exp \left(-\frac{1}{2} a_{-n}\left(Q_{nm} -\frac{Q_n Q_m}{ Q} \right)
  a_{-m} -c_{-n} R_{nm} b_{-m}
\right)| 0 ; 0\rangle
\label{eq:s2}
\end{equation}

This diagram represents a tadpole in the D25-brane background.  A
thorough analysis of this diagram should give a number of interesting
results.  We restrict ourselves here to some brief comments on this
subject, however.  Just as for the tree-level 4-point amplitude
discussed above, the integrand in (\ref{eq:s2}) converges rapidly for
$x < 1$ in the oscillator level-truncation approximation.  The
integrand has singularities at $x = 0$ and $x = 1$, corresponding to
the closed and open string tachyons.  At $x = 1$, the quadratic term
in the exponential has a limit where $R_{nm} = C_{nm}$, with a similar
result for the matter fields.  This form of squeezed state was
identified by Shapiro and Thorn as being an open string representation
of a closed string state \cite{Shapiro-Thorn}.  With a more careful
analysis it should be possible to show that this diagram gives rise to
precisely the closed string tadpoles expected from the bosonic disk
diagram with a single closed string vertex operator at a point in the
interior.  This result would demonstrate that when quantum effects are
included, the open string field is naturally pushed in a direction
associated with turning on closed string degrees of freedom, without
having to introduce the closed strings by hand.

Related issues to those just mentioned arise when one considers the
one-loop nonplanar 2-point function, which has a similar expression to
(\ref{eq:s2}).  This amplitude was computed in string field theory
using the conformal mapping method by Freedman, Giddings, Shapiro and
Thorn \cite{fgst}, who showed that the closed string poles naturally
appear in this amplitude and are associated with states of the
Shapiro-Thorn form mentioned above.  It would be very satisfying to
see how this result arises from the formalism of this paper.  Further
study of one-loop amplitudes in the open bosonic string using these
methods is left to future work.

\subsection{Higher loops}
\FIGURE{
\epsfig{file=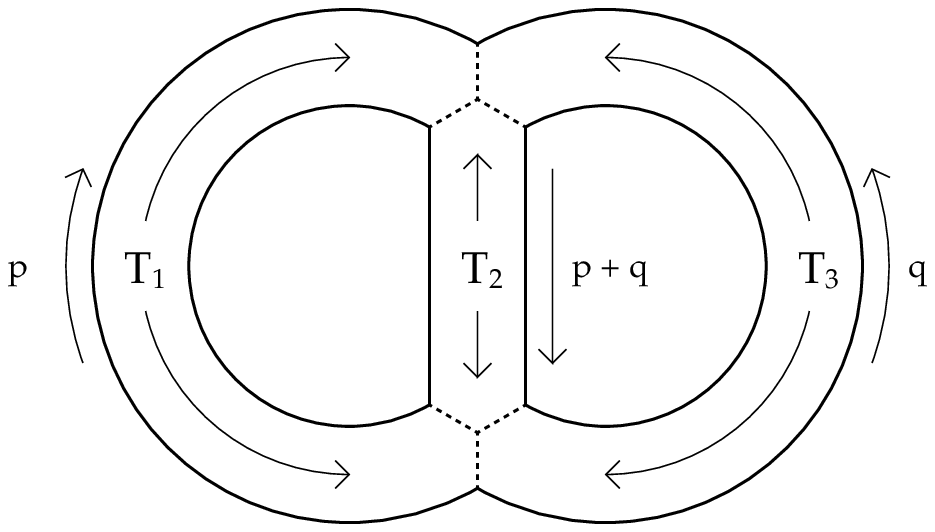,width=15cm}
\caption{\footnotesize Two-loop vacuum graph}
\label{f:2-loops}
}

It is clear that the methods we have developed here can be generalized
to diagrams with an arbitrary number of internal loops.  As a final
example, let us consider the simplest 2-loop diagram,
Figure~\ref{f:2-loops}, giving a contribution to the vacuum energy.
This diagram gives a contribution of
\begin{equation}
{\cal F}_{\rm 2-loop} =
\int \frac{dx}{x^2}  \frac{dy}{ y^2}  \frac{dz}{z^2}  \;
d^{26} p\,d^{26} q \;
\frac{\det (1+S \tilde{X})}{ \det (1-S \tilde{N})^{13}}
\exp \left( -\frac{1}{2}\left( q \; p \right) Q \left(\begin{array}{c}
q\\p
\end{array}\right) \right)
\label{eq:2l}
\end{equation}
where
\begin{equation}
x = e^{-T_1}, \;\;\;\;\; 
y = e^{-T_2}, \;\;\;\;\; 
z = e^{-T_3},
\end{equation}
\begin{equation}
S =
\left(\begin{array}{cccccc}
0 & 0 & 0 & C & 0 & 0\\
0 & 0 & 0 & 0 & 0 & C\\
0 & 0 & 0 & 0 & C & 0\\
C & 0 & 0 & 0 & 0 & 0\\
0 & 0 & C & 0 & 0 & 0\\
0 & C & 0 & 0 & 0 & 0\\
\end{array} \right),
\end{equation}
\begin{equation}
\tilde{N} =
\left(\begin{array}{cccccc}
\hat{N}^{11} (x, x) & \hat{N}^{12} (x, z) & \hat{N}^{13} (x, y) & 0 & 0 & 0\\
\hat{N}^{21} (z, x) & \hat{N}^{22} (z, z) & \hat{N}^{23} (z, y) & 0 & 0 & 0\\
\hat{N}^{31} (y, x) & \hat{N}^{32} (y, z) & \hat{N}^{33} (y,  y) & 0 & 0 & 0\\
0 & 0 & 0&\hat{N}^{11} (x, x) & \hat{N}^{12} (x, y) & \hat{N}^{13} (x, z) \\
0 & 0 & 0&\hat{N}^{21} (y, x) & \hat{N}^{22} (y, y) & \hat{N}^{23} (y, z) \\
0 & 0 & 0&\hat{N}^{31} (z, x) & \hat{N}^{32} (z, y) & \hat{N}^{33} (z, z) 
\end{array} \right),
\label{eq:2ln}
\end{equation}
and where $\tilde{X}$ is given by an identical expression to
(\ref{eq:2ln}), but with $\hat{N}$ replaced by $\hat{X}$.  The
momentum matrix $Q$ from (\ref{eq:2l}) is given by
\begin{equation}
Q = 4 N^{11}_{00}
 \left(\begin{array}{cc}
1 & 1\\
1 & 1\\
\end{array}\right)
+ \left(\begin{array}{cc}
2T_2 + 2T_3 & T_2\\
T_2&2T_1 + 2T_2
\end{array}\right)
+D^T \frac{1}{1-S \tilde{N}}  SD
\end{equation}
where
\begin{equation}
D = \left(\begin{array}{cc}
\;\;\;-\hat{N}^{12}_{n0} (x, 1) + \hat{N}^{13}_{n0} (x, 1) \;\;\; &
\;\;\;\hat{N}^{11}_{n0} (x, 1) - \hat{N}^{13}_{n0} (x, 1) \;\;\; \\
\;\;\;-\hat{N}^{22}_{n0} (z, 1) + \hat{N}^{23}_{n0} (z, 1) \;\;\; &
\;\;\;\hat{N}^{21}_{n0} (z, 1) - \hat{N}^{23}_{n0} (z, 1) \;\;\; \\
\;\;\;-\hat{N}^{32}_{n0} (y, 1) + \hat{N}^{33}_{n0} (y, 1) \;\;\; &
\;\;\;\hat{N}^{31}_{n0} (y, 1) - \hat{N}^{33}_{n0} (y, 1) \;\;\; \\
\;\;\;\hat{N}^{12}_{n0} (x, 1) - \hat{N}^{13}_{n0} (x, 1) \;\;\; &
\;\;\;-\hat{N}^{11}_{n0} (x, 1) + \hat{N}^{13}_{n0} (x, 1) \;\;\; \\
\;\;\;\hat{N}^{22}_{n0} (y, 1) - \hat{N}^{23}_{n0} (y, 1) \;\;\; &
\;\;\;-\hat{N}^{21}_{n0} (y, 1) + \hat{N}^{23}_{n0} (y, 1) \;\;\; \\
\;\;\;\hat{N}^{32}_{n0} (z, 1) - \hat{N}^{33}_{n0} (z, 1) \;\;\; &
\;\;\;-\hat{N}^{31}_{n0} (z, 1) + \hat{N}^{33}_{n0} (z, 1) \;\;\; \,.
\end{array}\right)
\end{equation}
Integrating over $p, q$ gives
\begin{equation}
{\cal F}_{\rm 2-loop} =
\int \frac{dx}{x^2}  \frac{dy}{ y^2}  \frac{dz}{z^2}  \;
f (x, y, z),
\end{equation}
where
\begin{equation}
f (x, y, z) =
\frac{\det (1+S \tilde{X})}{ [\det (1-S \tilde{N}) \det Q]^{13} }
\end{equation}
This gives the 2-loop contribution to the vacuum energy of open string
field theory in the standard vacuum containing a D25-brane.  This
expression contains divergences, such as that arising from the closed
string tachyon when the parameters $x, y$ and $z$ go to $1$, so the
integral is not finite.  At generic values of the parameters, however,
the integrand is finite.  A level-truncation analysis of the integrand
at fixed values of $x, y, z$, such as at $x = y = z = 1/2$, indicates
that the value of the integrand converges exponentially rapidly as
higher level oscillators are included, just as for the integrand of
the four-point tree amplitude and the one-point one-loop amplitude
described above.  Further study of loop amplitudes is left to future
work.  It would be particularly interesting to study the convergence
properties of the integrand for nonplanar diagrams at higher genus.

\section{Discussion}

In this paper we have presented a simple algorithm which gives a
closed-form expression for any open string amplitude at any loop
order, using string field theory.  For any diagram, the resulting
amplitude is an integral over a finite number of well-defined modular
parameters of a function of some infinite-dimensional matrices built
from a finite number of blocks containing the Witten 3-string vertex.
Using level truncation on oscillator level, this gives an algorithm
for systematically computing any open string amplitude to an arbitrary
degree of accuracy.  We find that for both tree and loop diagrams, the
integrand of the amplitude converges very rapidly under level
truncation, although the rate of convergence becomes slower as the
Witten parameter $T$ associated with the length of an internal edge of
a diagram goes to 0.  For all finite tree amplitudes we considered, we
found that the error introduced by a truncation at oscillator level
$L$ goes as $1/L$.

There are many potentially interesting applications of this approach,
and many ways in which it would be interesting to extend this work.
We mention a few of these directions briefly here.
\vspace{0.05in}

\noindent$\bullet$ 
We do not yet have sufficient analytic control of
the infinite matrices appearing in the string field theory amplitudes
to evaluate even the simplest diagrams exactly.  It would be very
interesting to generalize the recent work on the diagonalization of
the matrices of Neumann coefficients for the Witten vertex
\cite{diagonalization} to the matrices $\hat{N}$ defined through
(\ref{eq:nti}).  Such a generalization might lead naturally to exact
analytic expressions for amplitudes, at least for simple diagrams.
\vspace{0.03in}

\noindent$\bullet$ 
We find empirically that truncation at
oscillator level $L$ of the string field theory expression for the
Veneziano amplitude and its off-shell generalization introduces an
error of order $1/L$.  We found the same kind of convergence for the
off-shell $p = 0$ 5-tachyon amplitude.  It would be interesting to
prove this result rigorously for these amplitudes, and to understand
the rate of convergence of this approximation for more complicated
diagrams.
\vspace{0.03in}

\noindent$\bullet$ While the analysis of this paper focuses on
Witten's cubic open bosonic string field theory, the methods can be
generalized to other string field theories.  It would be particularly
interesting to apply this method to superstring field theory, either
in the Berkovits \cite{Berkovits} or Witten
\cite{Witten-super,Arefeva} formulation.  Since there is currently no
method available for explicitly computing covariant on-shell
correlation functions of superstrings at high loop order/genus (see
\cite{dp} for some recent work at two-loop order in closed strings),
this method might lead to interesting new results for on-shell
superstring amplitudes.  This method can also be used as a means of
checking the various proposals for superstring field theory.  Since
loop diagrams in superstring field theory do not have the divergence
problems arising from tachyons which afflict loop diagrams in the
bosonic theory, we expect that this approach may provide a useful
means of numerically calculating high-order loop diagrams in
superstring theory.
\vspace{0.03in}

\noindent$\bullet$
A crucial question in open string field theory, which has been brought
back to the fore by recent developments related to Sen's tachyon
condensation conjectures, is how closed strings are encoded in open
string field theory, and whether closed strings can be treated as
asymptotic on-shell states using only the degrees of freedom of open
string field theory \cite{closed}.  A careful study of open string
loop amplitudes using the methods of this paper may lead to new
insight into this question.
\vspace{0.03in}

\noindent$\bullet$ Related to the closed string issue is that of what
the natural degrees of freedom are which should describe open string
field theory after the tachyon has condensed into the stable vacuum.
In this vacuum there are no open string degrees of freedom
\cite{Ellwood-Taylor,efhm}.  It would
be very interesting if the methods of this paper could be generalized
to study the vacuum string field theory (VSFT) in this stable vacuum,
either directly through expanding around the nontrivial solution of
the Witten theory as in \cite{Ellwood-Taylor,efhm}, or using the RSZ
pure ghost Ansatz for the BRST operator \cite{RSZ-VSFT}
\vspace{0.03in}

\noindent$\bullet$ Finally, the methods described here might be used
to compute the effective action for the gauge field and/or the tachyon
on a system of multiple D$p$-branes, along the lines of \cite{WT-SFT}.
This might lead to new insight into the form of the resulting
nonabelian Born-Infeld action and/or the tachyon effective field
theory, neither of which are yet completely understood.

\section*{Acknowledgements}

I would like to thank Erasmo Coletti, Ian Ellwood, Steve Giddings,
Michael Green, Ashoke Sen, Jessie Shelton, Ilia Sigalov, and Barton
Zwiebach for helpful discussions.  I would also like to thank the
Newton Institute and the organizers of the Newton Institute workshop
on M-theory for support and hospitality while much of this work was
done.  The numerical computations described in this work were done
using {\it Mathematica}.
This work was supported by the DOE through contract
\#DE-FC02-94ER40818.

\newpage
\appendix

\section{Analytic description of 4-tachyon amplitude}

In this Appendix we briefly summarize the results of
\cite{Giddings,Sloan,Samuel-off} which give an analytic description of the
on-shell and off-shell 4-tachyon amplitude from string field theory.

In \cite{Giddings}, Giddings gave an explicit conformal map which
takes the Riemann surface defined by the Witten diagram with internal
edge of length $T$ to the standard disc with four tachyon vertex
operators on the boundary, parameterized by the Koba-Nielsen parameter
$\xi$ (usually called $x$).  This conformal map is defined in terms of
four parameters $\alpha, \beta, \gamma, \delta$ satisfying the
relations
\begin{equation}
\alpha \beta = \gamma \delta = 1
\label{eq:relation1}
\end{equation}
and
\begin{equation}
\frac{1}{2}= \Lambda_0 (\theta_1, k) -\Lambda_0 (\theta_2, k)
\label{eq:relation2}
\end{equation}
where $\Lambda_0 (\theta, k)$ is Heuman's lambda function, defined as
\begin{equation}
\Lambda_0 (\theta, k) = \frac{2}{ \pi} 
\left( E (k) F (\theta, k') + K (k)
(E (\theta, k') -F (\theta, k')) \right)
\end{equation}
in terms of the complete elliptic integrals of the first and second
kinds $K (k), E (k)$ and the incomplete elliptic integral of the first
kind $F (\theta, k)$, and the parameters $\theta_1, \theta_2, k, k'$
given by
\begin{eqnarray}
k = \gamma^2 & \hspace{1in} &  k' = \sqrt{1-k^2}\\
 \theta_1 = \sin^{-1} \frac{\beta}{ \sqrt{\beta^2 + \gamma^2}}  && 
 \theta_2 = \sin^{-1} \frac{\alpha}{ \sqrt{\alpha^2 + \gamma^2}}  \,.
\end{eqnarray}
Since the four parameters $\alpha-\delta$ are related by the three
relations (\ref{eq:relation1}, \ref{eq:relation2}), all the parameters
can be considered as functions of $\alpha$.

The Witten parameter $T$ is related to the parameter $\alpha$ through
\begin{equation}
T = 2K (k') \left( Z (\theta_2, k') -Z (\theta_1, k') \right)
\end{equation}
where
$Z (\theta, k)$ is Jacobi's zeta function
\begin{equation}
Z (\theta, k) = E (\theta, k) -\frac{E (k)}{ K (k)}  F (\theta, k)\,.
\end{equation}
The parameter $\alpha$ is in turn related to the Koba-Nielsen
parameter $\xi$ through
\begin{equation}
\alpha = \sqrt{\frac{1-\sqrt{\xi}}{1 + \sqrt{ \xi}} }\,.
\end{equation}
When $T = 0$ we have $\alpha = -1 + \sqrt{2}$ and $\xi = 1/2$.  When $T
= \infty$ we have $\alpha = 0$ and $ \xi = 1$.  For $\xi < 1/2$, the
associated Witten diagram is in the opposite (t) channel from that
shown in Figure~\ref{f:diagram1}.  This explains the two terms in
(\ref{eq:4t1}).

In \cite{Giddings}, Giddings used the conformal map just described to
map the string field theory calculation of the on-shell 4-tachyon
amplitude to a conformal field theory calculation, and showed that the
result of the string field theory calculation is indeed the Veneziano
amplitude.  In \cite{Sloan,Samuel-off}, Samuel and Sloan used Giddings'
approach and with some additional analysis found an analytic formula
for the off-shell 4-tachyon amplitude.  The off-shell amplitude
differs from the Veneziano amplitude in that it has an additional term
in the integrand of (for $1/2 < \xi < 1$, using the notation of
\cite{Samuel-off})
\begin{equation}
\left(\frac{ \kappa (\xi)}{ 2}  \right)^{p_1^2 + p_2^2 + p_3^2 + p_4^2
-4}
\label{eq:off-shell-correction}
\end{equation}
where
\begin{equation}
\kappa = \exp \left( -N \beta \int_1^\infty dw \; \ln (w-1)
\frac{{\rm d}}{ {\rm d} w}  \left[ \frac{\sqrt{(w^2 + a^2 \gamma^2)
(w^2 + \alpha^2 \delta^2)}}{ (w + 1) (\beta^2 w^2 -\alpha^2)}  \right]\right)
\label{eq:kappa}
\end{equation}
with
\begin{equation}
N = 2 \alpha \frac{\beta^2 -\alpha^2}{ \sqrt{(\alpha^2 + \gamma^2)
(\alpha^2 + \delta^2)}} \,.
\end{equation}
It is clear that (\ref{eq:off-shell-correction}) vanishes on-shell,
where $p_i^2 = 1$.  

Relating  the off-shell Veneziano integrand at $p = 0$ with the
correction (\ref{eq:off-shell-correction}) to the integrand appearing
in (\ref{eq:4tap}), the expressions agree if \cite{Samuel-off}
\begin{equation}
\frac{3^9}{2^{12} x}
\frac{\det (1+S \tilde{X})}{\det (1-S \tilde{N})^{13}} 
=  \frac{4}{\pi \kappa^4}  \sqrt{(\alpha^2 + \gamma^2) (\beta^2 +
\gamma^2)} (\beta^2 -\alpha^2) K (\gamma^2)\,.
\label{eq:analytic-integrand}
\end{equation}
The extent to which these expressions agree at a finite level of
truncation controls the rate of convergence of the level-truncated
approximations to the on-shell and off-shell 4-point amplitudes, such
as those discussed in Section 2.1 and 2.2.  In
Figure~\ref{f:graph-c4i} we have graphed the RHS of
(\ref{eq:analytic-integrand}) and compared with low-level truncations.
As discussed in Section 2.1, the convergence is extremely fast except
near $x = 1$.

The comparison just done relates the analytic formula for the
momentum-independent part of the off-shell four-tachyon amplitude
found in \cite{Sloan,Samuel-off} to the matrix expression calculated using
the oscillator methods of this paper.  It is also possible to relate
the momentum-dependent factors.  In \cite{Sloan,Samuel-off}, explicit
formulae were given for the momentum-dependent factors in the
off-shell extension of the Veneziano integrand.  These can be related
to the matrices $Q_{ij}$ of (\ref{eq:4tq}), although the comparison is
not direct since the matrices are subject to a redefinition using the
conservation law $\sum_{i} p_i = 0$.  The computation of the on-shell
Veneziano amplitude in Section 2.2 demonstrates that the
momentum-dependent terms also agree on shell.

\normalsize

\bibliographystyle{plain}

\end{document}